# An *ab initio* dataset of size-dependent effective thermal conductivity for advanced technology transistors

Han Xie(谢涵)[1,2], Ru Jia(贾如)[3], Yonglin Xia(夏涌林)[3], Lei Li(李磊)[1], Yue Hu(胡跃)[4], Jiaxuan Xu(徐家璇)[3], Yufei Sheng(盛宇飞)[3], Yuanyuan Wang(王元元)[1,5,*], Hua Bao(鲍华)[6,*]

[1]School of Energy and Materials, Shanghai Polytechnic University, Shanghai 201209, China.

[2]Institute of Integrated Circuits, Shanghai Polytechnic University, Shanghai 201209, China.

[3]University of Michigan-Shanghai Jiao Tong University Joint Institute, Shanghai Jiao Tong University, Shanghai 200240, China.

[4]CTG Wuhan Science and Technology Innovation Park, China Three Gorges Corporation, Wuhan 430010, China.

[5]Shanghai Thermophysical Properties Big Data Professional Technical Service Platform, Shanghai Polytechnic University, Shanghai 201209, China.

[6]Global Institute of Future Technology, Shanghai Jiao Tong University, Shanghai 200240, China.

## Abstract

As the size of transistors shrinks and power density increases, thermal simulation has become an indispensable part of the device design procedure. However, existing works for advanced technology transistors use simplified empirical models to calculate effective thermal conductivity in the simulations. In this work, we present a dataset of size-dependent effective thermal conductivity with electron and phonon properties extracted from *ab initio* computations. Absolute in-plane and cross-plane thermal conductivity data of eight semiconducting materials (Si, Ge, GaN, AlN, 4H-SiC, GaAs, InAs, BAs) and four metallic materials (Al, W, TiN, Ti) with the characteristic length ranging from 5 to 50 nanometers have been provided. Besides the absolute value, normalized effective thermal conductivity is also given, in case it needs to be used with updated bulk thermal conductivity in the future. The dataset presented in this paper are

---

* Authors to whom correspondence should be addressed. Electronic addresses: wangyuanyuan@sspu.edu.cn and hua.bao@sjtu.edu.cn

## 1. Introduction

As the size of transistors used in integrated circuits shrinks and power density increases, heat dissipation has become a critical issue[1]. As a result, thermal simulation has become an indispensable part of the device design procedure, where Fourier's law is used[2]. However, size effects in the ballistic-diffusive regime renders Fourier's law invalid at micro/nanoscales[3–8]. Consequently, employing size-dependent effective thermal conductivity for the nanostructures in electronic devices is a mainstream strategy[5,9,10].

In advanced technology transistors, such a strategy has been widely used to study the thermal transport in nanosheets[2,10–12] and nanowires[13–15]. Most of these works focus on the self-heating effect in transistors but use simplified empirical models to extract the thermal conductivity. For example, Ma *et al.*[10] used a Debye-Callaway model to predict the size-dependent in-plane and cross-plane effective thermal conductivity of GaN thin films. This empirical model is derived under the assumption that the dispersion curve of acoustic phonons is linearly related to phonon wavevectors[16,17], which will render the computed thermal conductivity inaccurate. Venkateswarlu *et al.*[12] employed a Connelly thermal conductivity model[18,19] to obtain the thermal conductivity of different regions. However, that model consists of fitting parameters and cannot capture the actual physical mechanisms behind thermal transport. The disadvantage in accuracy and lack of predictive power due to fitting parameters in these empirical models necessitates a parameter-free and in-depth study of thermal transport in nanostructures.

From the experimental point of view, researchers have devoted to measure the size-dependent thermal conductivity and study the physical mechanisms[20–22]. Li *et al.*[20] investigated the size effect on the thermal conductivity of graphene oxide nanosheets. Larger-sized nanosheets were found to have higher thermal conductivity, which was explained by the fewer morphological defects rather than structural defects. Jaffe *et al.*[21] measured the cross-plane thermal conductivity of hexagonal boron nitride and explored the thickness dependence. The thermal conductivity was found to increase with the increasing thickness and significant contribution from phonons with mean free paths on the order of several hundred nanometers were observed for thicker flakes. Ziade *et al.*[22] measured the size-dependent thermal conductivity of GaN at different temperatures with the frequency domain themoreflectance (FDTR) technique. The decrease of thermal conductivity in GaN films with smaller thickness was explained by the ballistic phonon transport under 150 nm. These experimental measurements[20–22] of size-dependent thermal conductivity can capture all the affecting factors and give reasonably good results. However, the reported value is often for only one material in each work due to the complexity of experimental setup.

The above-mentioned works either exhibit the theoretical classical behavior with inaccuracies or have the limitation that only one material is studied experimentally each time. Recently, Wang and Bao[23] studied the size-dependent thermal conductivity of nine metallic nanowires. Their work can give an important guidance in selecting advanced interconnects but nanosheets are not considered. In this work, we combined parameter-free *ab initio* computations and analytical equation to obtain the size-dependent effective thermal conductivity of nanosheets. A dataset containing eight semiconducting materials (Si, Ge, GaN, AlN, 4H-SiC, GaAs, InAs, BAs) and four metallic materials (Al, W, TiN, Ti) representative for advanced technology transistors is presented. Size-dependent in-plane and cross-plane effective thermal conductivity values are provided and the underlying physical mechanisms are discussed.

## 2. Methods

To obtain the size-dependent effective thermal conductivity, the bulk thermal conductivity and thermal properties of heat carriers are generally computed as the first step. Size effects are then introduced by either the suppression function or Boltzmann transport equation (BTE).

**2.1 Bulk thermal conductivity**

In semiconductors, phonons are the primary heat carriers, while in metals, both electrons and phonons significantly contribute to heat conduction[24,25]. For isotropic

materials, phonon thermal conductivity in bulk materials can be expressed as[26]

$$k_{\text{ph}} = \frac{1}{3} \sum_{\lambda} c_\lambda v_\lambda^2 \tau_\lambda, \qquad (1)$$

where $c_\lambda$, $v_\lambda$, and $\tau_\lambda$ are the volumetric heat capacity, group velocity and relaxation time of phonons, respectively. $\lambda = (\boldsymbol{q}, v)$ represent different phonon modes that can be distinguished by wavevector $\boldsymbol{q}$ and phonon branch $v$. Electron thermal conductivity can be obtained with a similar equation:

$$k_{\text{el}} = \frac{1}{3} n_s \sum_{n\boldsymbol{k}} c_{n\boldsymbol{k}} v_{n\boldsymbol{k}}^2 \tau_{n\boldsymbol{k}}, \qquad (2)$$

where $c_{n\boldsymbol{k}}$, $v_{n\boldsymbol{k}}$, and $\tau_{n\boldsymbol{k}}$ are the volumetric heat capacity, group velocity and relaxation time of electrons, respectively. The summation is over all the electron states specified by band index $n$ and wavevector $\boldsymbol{k}$. $n_s$ denotes the number of particles per state, which will be taken as 2 to account for spin degeneracy. If spin degeneracy was already considered in *ab initio* computations, $n_s = 1$ will be used. The bulk thermal conductivity can be calculated with

$$k_{\text{bulk}} = \begin{cases} k_{\text{ph}}, & \text{for semiconductors.} \\ k_{\text{ph}} + k_{\text{el}}, & \text{for metals.} \end{cases} \qquad (3)$$

Regarding the relaxation time, phonon-phonon and phonon-isotope scattering were considered in our computation for semiconductors. For metals, phonon-phonon scattering, phonon-electron scattering and phonon-isotope scattering were considered for $\tau_\lambda$, while only electron-phonon scattering was considered for $\tau_{n\boldsymbol{k}}$.

**2.2 Size-dependent effective thermal conductivity**

With the mode-level electron and phonon properties as the input, size-dependent effective thermal conductivity can be computed from either analytical equation with suppression function or BTE. With analytical equation, the in-plane and cross-plane effective thermal conductivity for semiconductors can be computed with

$$k_{\text{in}} = \frac{1}{3} \sum_{\lambda} c_\lambda v_\lambda^2 \tau_\lambda S_{\text{in}}(\lambda), \qquad (4)$$

and

$$k_{\text{cross}} = \frac{1}{3} \sum_{\lambda} c_\lambda v_\lambda^2 \tau_\lambda S_{\text{cross}}(\lambda), \qquad (5)$$

where $S_{\text{in}}(\lambda)$ and $S_{\text{cross}}(\lambda)$ are suppression functions for phonons that define the ratio between suppressed mean free path and bulk mean free path[27,28].

For metals, the in-plane and cross-plane effective thermal conductivity can be computed with

$$k_{\text{in}} = \frac{1}{3}\sum_{\lambda} c_\lambda v_\lambda^2 \tau_\lambda S_{\text{in}}(\lambda) + \frac{1}{3}n_s \sum_{n\bm{k}} c_{n\bm{k}} v_{n\bm{k}}^2 \tau_{n\bm{k}} S_{\text{in}}(n\bm{k}), \quad (6)$$

and

$$k_{\text{cross}} = \frac{1}{3}\sum_{\lambda} c_\lambda v_\lambda^2 \tau_\lambda S_{\text{cross}}(\lambda) + \frac{1}{3}n_s \sum_{n\bm{k}} c_{n\bm{k}} v_{n\bm{k}}^2 \tau_{n\bm{k}} S_{\text{cross}}(n\bm{k}), \quad (7)$$

where $S_{\text{in}}(n\bm{k})$ and $S_{\text{cross}}(n\bm{k})$ are the same suppression functions for electrons with the same analytical forms as $S_{\text{in}}(\lambda)$ and $S_{\text{cross}}(\lambda)$. Details about the analytical form of these suppression functions can be found in the supplementary materials.

Besides the analytical method shown above, the effective thermal conductivity can also be obtained from solving the steady-state BTE shown as below[29]

$$\bm{v} \cdot \nabla e = -\frac{e - e^0}{\tau}, \quad (8)$$

where $\bm{v}$ is group velocity and $\tau$ is relaxation time. $e$ and $e^0$ are the energy density and energy density at equilibrium, respectively. GiftBTE was used to solve this equation for both electrons and phonons to extract size-dependent effective thermal conductivity.

## 2.3 Simulation details

The calculations in our dataset are performed using *ab initio* methods to obtain the mode-level electron and phonon properties. We start with calculating the electronic band structure using a self-consistent field (SCF) calculation following lattice optimization. The second-order interatomic force constants (2nd IFCs) are determined via density functional perturbation theory (DFPT), while third-order interatomic force constants (3rd IFCs) are computed using finite-difference supercell methods. The forces are derived from SCF calculations of various displaced supercell configurations. The THIRDORDER.PY package[30] is employed to construct the supercell for third-order IFCs. By combining the 2nd IFCs and 3rd IFCs, the ShengBTE package[30] is used to calculate phonon properties and lattice thermal conductivity. It is important to note that the parameters related to the supercell size and nearest neighbors are optimized until the lattice thermal conductivity converges. For electron-phonon (e-p) scattering, the e-p coupling matrix element is calculated using the electron-phonon Wannier (EPW) code[31]. This matrix element is initially computed on coarse grids for electron and phonon wavevectors and later interpolated onto denser grids using a maximally localized Wannier function basis[32]. An energy window of 1.5 eV is applied in the electron-phonon matrix calculations. When calculating the scattering rates, only irreducible points within the first Brillouin zone are considered, leveraging symmetry to significantly reduce computational costs.

## 3. Results and Discussion

### 3.1 Data validation

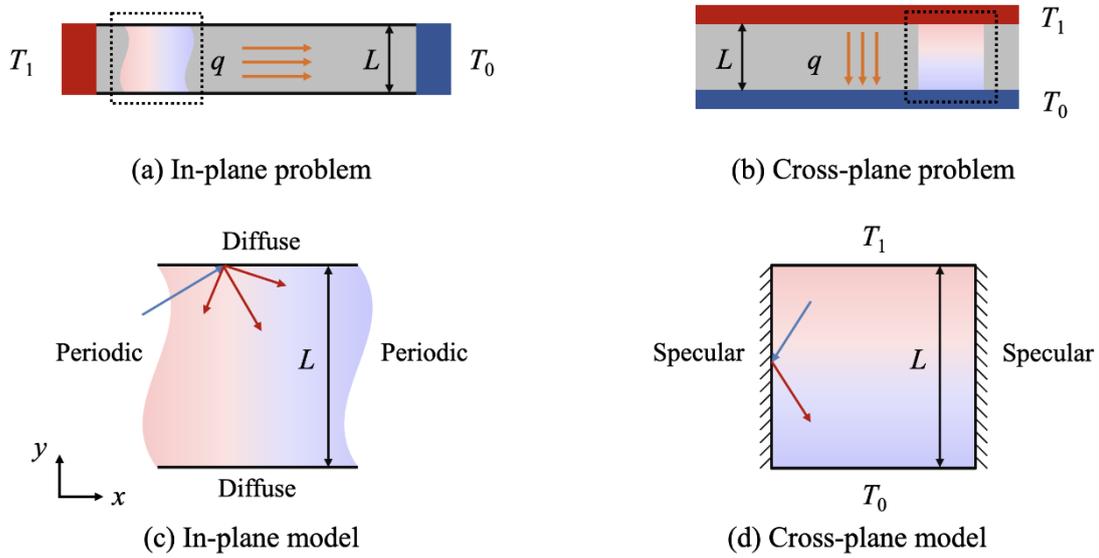

Fig. 1 Schematic for (a) in-plane and (b) cross-plane heat conduction problems in nanosheets, (c) in-plane and (d) cross-plane models used in GiftBTE computations

We have representatively selected eight semiconducting materials (Si, Ge, GaN, AlN, 4H-SiC, GaAs, InAs, BAs) and four metallic materials (Al, W, TiN, Ti) useful in advanced technology transistors and calculated their size-dependent effective thermal conductivity with the analytical method shown in Sec. 2. As shown in Fig. 1(a-b), in-plane thermal conductivity is defined as the ability to conduct heat parallel to the nanosheet plane while cross-plane thermal conductivity is defined based on the perpendicular direction. The effective thermal conductivity is defined by Fourier's law as $\kappa = -q/\nabla T$, where $q$ is the heat flux shown in Fig. 1(a-b) and $\nabla T$ is the temperature gradient.

To validate the correctness of our method, we have calculated the size-dependent thermal conductivity and compared the bulk thermal conductivity with experimental data. All our computation was carried out at the room temperature of 300 K. It can be seen from Fig. S1 in the supplementary materials that the effective thermal conductivity for both in-plane and cross-plane cases increase with the increasing size and converge to the bulk value. Since the EPW code can also apply spin degeneracy, we have carefully checked that the degeneracy was not considered redundantly in Eqns. (2) and (6-7). The computed bulk thermal conductivity results are compared with reference experimental values in Table 1. The good agreement between our computational results and experimental values proves the accuracy of our data. In advanced technology transistors, characteristic length of the typical nanosheet structures often lies in the range of 5 to 50 nm[11,12,33]. Therefore, we calculated the size-dependent in-plane and cross-plane effective thermal conductivity in this length range.

The data for semiconductors are illustrated as solid lines in Fig. 2 and those for metals are illustrated in Fig. 3. The circular dots in these two figures represent the effective thermal conductivity computed by numerically solving BTE. The models used in solving BTE with the GiftBTE package are shown in Fig. 1(c-d). In in-plane model, diffusive boundary conditions are used for top and bottom surfaces to represent the actual situation. Periodic boundary conditions are used for left and right surfaces because the thermal conductivity is a material property which is irrelevant of the large-enough size in $x$ direction. In cross-plane model, specular boundary conditions are used for left and right surfaces because phonons can be specularly reflected at these boundaries. It can be seen from Fig. 2 and Fig. 3 that the analytical results show good agreement with the GiftBTE results, which can further validate the correctness of our data.

Table 1 The computed bulk thermal conductivity in comparison with reference experimental values

| Material Type | Material Symbol | $\kappa_{bulk}$ (W/mK) | $\kappa_{reference}$ (W/mK) |
|---|---|---|---|
| Semiconductor | Si | 153.81 | 156[34] |
| | Ge | 54.36 | 60[34] |
| | GaN | 269.12 | 260[35,36] |
| | AlN | 295.85 | 319[37] |
| | 4H-SiC | 410.58 | 490[38] |

|  | GaAs | 42.21 | 45[39] |
|  | InAs | 33.39 | 27[40] |
|  | BAs | 1223.66 | 1300[41] |
| Metal | Al | 223.50 | 237[42] |
|  | W | 195.60 | 175[43,44] |
|  | TiN | 72.38 | 63[45] |
|  | Ti | 26.50 | 21[46] |

**3.2 Semiconductor data**

For semiconductors, only phonons contribute to the total thermal conductivity. From Fig. 2(a-b), we can find that BAs and AlN has very high in-plane and cross-plane effective thermal conductivity at 50 nm. It is necessary to clarify that four-phonon scattering process was considered for BAs due to the strong fourth-order anharmonicity[47–49]. For other materials, only three-phonon scattering process was considered. From Fig. S1, we can see that BAs has the highest bulk thermal conductivity, which can be explained by the large acoustic-optical energy gap[48,50,51] and small scattering rates. It is notable that BAs shows a sharper decreasing trend in effective thermal conductivity than other materials when the characteristic length decreases. At 5 nm, AlN shows the highest effective thermal conductivity for both in-plane and cross-plane cases. It can be found that materials with high bulk thermal conductivity does NOT necessarily have a high effective thermal conductivity at small scales. The difference between BAs and AlN can be explained by the large acoustic velocities of bulk AlN[52]. The shrinking of characteristic size mainly affects the phonon scattering rates instead of acoustic group velocities.

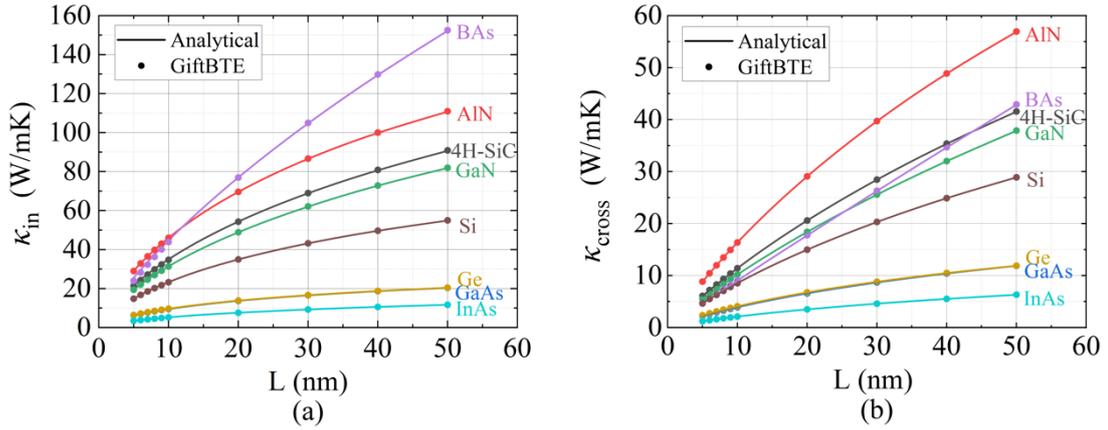

Fig. 2 Size-dependent (a) in-plane and (b) cross-plane effective thermal conductivity for semiconductors

### 3.3 Metal data

For metals, both electrons and phonons have a significant contribution to total thermal conductivity. Fig. 3(a-b) shows the in-plane and cross-plane effective thermal conductivity of four metallic materials. Since electrons have a significant contribution to the total thermal conductivity, considering third-order anharmonicity is enough to describe the phonon-phonon scattering processes in phonon contribution. It can be seen from Fig. 3 that Al shows the highest thermal conductivity over all the length range. Tungsten (W) has lower thermal conductivity than Al because it has less free electrons and has a larger contribution from phonons[53]. Ti has the lowest in-plane and cross-plane effective thermal conductivity over all the length range.

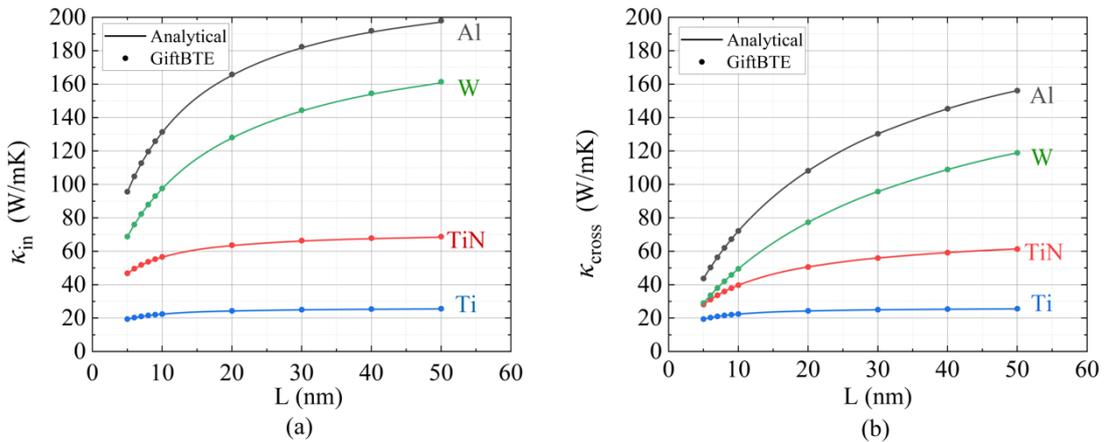

Fig. 3 Size-dependent (a) in-plane and (b) cross-plane effective thermal conductivity for metals

When computing cross-plane effective thermal conductivity, directional discretization[29] with $N_\theta = 4$ and $N_\varphi = 4$ and spatial discretization with 100×100 mesh was used after convergence test, as shown in the supplementary materials. For in-plane effective thermal conductivity, denser directional discretization[29] with $N_\theta = 16$ and $N_\varphi = 16$ is used for $L$ from 5 to 50 nm and spatial discretization with 10×10 mesh was used after convergence test. In both Fig. 2 and Fig. 3, we can see that cross-plane thermal conductivity is smaller than in-plane thermal conductivity. Similar phenomenon is also observed for GaN[10]. Since the suppression mechanism for cross-plane thermal conductivity is mainly phonon ballistic transport leading to boundary temperature jump and the mechanism for in-plane thermal conductivity is mainly phonon boundary scattering resulting in boundary heat flow slip[9]. This indicates that ballistic transport has a relatively stronger suppression effect than boundary scattering at the same length scale.

### 3.4 Normalized effective thermal conductivity

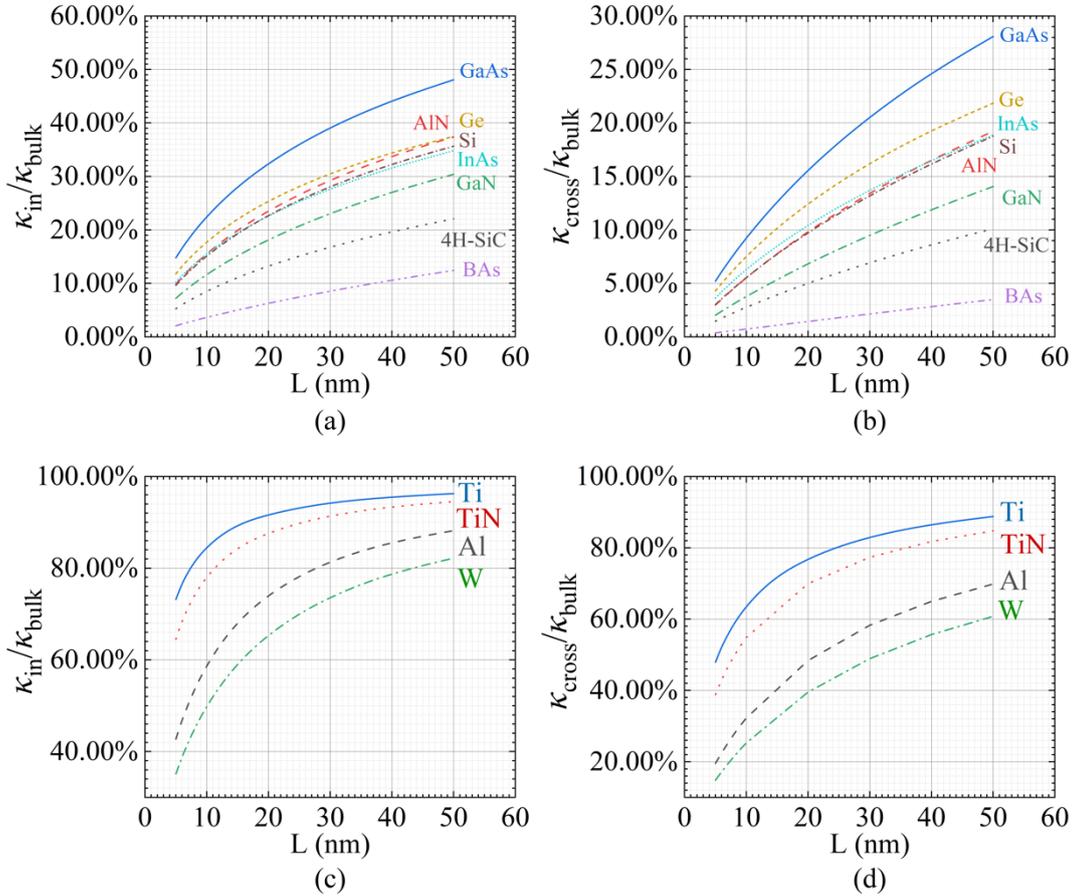

Fig. 4 Size-dependent (a) normalized in-plane effective thermal conductivity for semiconductors, (b) normalized cross-plane effective thermal conductivity for semiconductors, (c) normalized in-plane effective thermal conductivity for metals, (d) normalized cross-plane effective thermal conductivity for metals

In case the bulk thermal conductivity needs to be updated in the future, we have also provided the normalized effective thermal conductivity and illustrated the data in Fig. 4. Fig. 4(a-b) shows the normalized in-plane and cross-plane effective thermal conductivity for semiconductors and Fig. 4(c-d) shows the normalized thermal conductivity for metals. Grid lines are plotted in these figures to help in finding the data. Our intended users can find the normalized thermal conductivity according to the characteristic length from Fig. 4 and the bulk thermal conductivity from Table 1. We have also uploaded our data to Science Data Bank[54]. The multiplication of these two values yields the effective in-plane/cross-plane thermal conductivity.

## 4. Conclusion

In summary, we have calculated the size-dependent in-plane and cross-plane effective thermal conductivity from *ab initio* computations. Eight semiconducting materials (Si, Ge, GaN, AlN, 4H-SiC, GaAs, InAs, BAs) and four metallic materials (Al, W, TiN, Ti) were chosen to present the data. The accuracy and correctness of our data are proved by comparing calculated bulk thermal conductivity with experimental data and comparing effective size-dependent thermal conductivity computed from analytical method with GiftBTE result. A dataset containing the absolute value for effective thermal conductivity as well as the normalized value is provided. The data for characteristic length ranging from 5 to 50 nanometers is given in Figs. (2-4) and in csv format at https://doi.org/10.57760/sciencedb.j00113.00154. Within the isotropic assumption, these data can be used directly as input for finite element methods (FEM) in thermal simulation. Please cite our paper when you use our data. This work may shed some light on the thermal management of advanced technology transistor design.

## Acknowledgement

This work is supported by the National Key R&D Project from Ministry of Science and Technology of China (Grant No. 2022YFA1203100), the National Natural Science Foundation of China (Grant No. 52122606) and the funding from Shanghai Polytechnic

University. We would like to thank Jiayang Hui and Xinrui Wang from Shanghai Polytechnical University for their help in organizing some data. The computations in this paper were run on the π 2.0 cluster supported by the Center for High Performance Computing at Shanghai Jiao Tong University.